\documentclass[journal]{IEEEtran}
\ifCLASSINFOpdf
 \else
 \fi

\usepackage{cite}
\usepackage{amsmath,amssymb,amsfonts}

\DeclareMathOperator*{\argmin}{arg\,min}
\usepackage{graphicx}
\usepackage{textcomp}
\usepackage{xcolor}
\usepackage{subfigure}
\usepackage[english]{babel}
\usepackage{tabularx}
\usepackage{algorithm}
\usepackage{algpseudocode}
\usepackage{caption}
\usepackage{subcaption}

\usepackage{amsthm} 
\theoremstyle{plain}
\newtheorem{theorem}{Theorem} 
\theoremstyle{definition}
\newtheorem{assumption}{Assumption}
\newtheorem{definition}{Definition}
\theoremstyle{remark}
\newtheorem{remark}{Remark}

\begin{document}
%
\title{{Deep Neural Network-Based Aerial Transport in the Presence of Cooperative and Uncooperative UAS}}
%
%
%

\author{Muhammad Junayed Hasan Zahed and Hossein Rastgoftar
\thanks{Muhammad Junayed Hasan Zahed is with the Department of Aerospace and Mechanical Engineering (AME), University of Arizona, Tucson, AZ, USA, Email: mjhz@arizona.edu.}%

\thanks{Hossein Rastgoftar is with the Aerospace and Mechanical Engineering (AME) and Electrical and Computer Engineering (ECE) Departments, University of Arizona, Tucson, AZ, USA, Email: hrastgoftar@arizona.edu.}
}
\maketitle


\begin{abstract}
We present a resilient deep neural network (DNN) framework for decentralized transport and coverage using uncrewed aerial systems (UAS) operating in 
$\mathbb{R}^n$. The proposed DNN-based mass-transport architecture constructs a layered inter-UAS communication graph from an initial formation, assigns time-varying communication weights through a forward scheduling mechanism that guides the team from the initial to the final configuration, and ensures stability and convergence of the resulting multi-agent transport dynamics. The framework is explicitly designed to remain robust in the presence of uncooperative agents that deviate from or refuse to follow the prescribed protocol. Our method preserves a fixed feed-forward topology but dynamically prunes edges to uncooperative agents, maintains convex, feedforward mentoring among cooperative agents, and computes global desired set points through a sparse linear relation consistent with leader references. The target set is abstracted by $N$ points that become final desired positions, enabling coverage-optimal transport while keeping computation low and guarantees intact. Extensive simulations demonstrate that, under full cooperation, all agents converge rapidly to the target zone with a 10\% boundary margin and under partial cooperation with uncooperative agents, the system maintains high convergence among cooperative agents with performance degradation localized near the disruptions, evidencing graceful resilience and scalability.  These results confirm that forward-weight scheduling, hierarchical mentor--mentee coordination, and on-the-fly DNN restructuring yield robust, provably stable UAS transport in realistic fault scenarios.
\end{abstract}
\begin{IEEEkeywords}
Deep Neural Network-Based UAS Transport, Cooperative  agents, Uncooperative agents, Target Coverage
\end{IEEEkeywords}

\section{Introduction}

 Mass transport is widely studied as the transformation of a collection of masses, agents, or particles from one distribution to another under a prescribed cost. Beyond theory, Optimal Mass Transport (OMT)-style transport has powered applications in brain morphometry and neuroimaging \cite{GERBER2023102696, ma2019optimal, 7053911}, cancer histopathology \cite{wang2013linear}, 3D tumor segmentation \cite{lin20213d}, image processing and machine learning \cite{shakib2020mass, 7974883, song2023chest}, domain adaptation \cite{courty2017joint}, inverse problems \cite{doi:10.1137/19M1261122} and quantitative medical imaging inversion \cite{WU2025107505} and so on. 

In this work, we recast mass transport as a safe, decentralized multi-agent coordination problem: drive a team from an arbitrary initial formation to a distributed target configuration, subject to time-varying communication graphs and keep-in/keep-out constraints. Building on our DNN-based transport architecture \cite{11007873}, we present a fault-resilient, assignment-free method that (i) fixes a layered feed-forward inter-UAS graph synthesized from the initial formation, (ii) schedules time-varying convex weights forward in time, and (iii) prunes edges around uncooperative agents while preserving convex-hull containment and stability.

\subsection{Related work}

Dynamic and computational OMT furnish geometry-aware metrics and flows for morphing densities \cite{benamou2000computational,cuturi2013sinkhorn}. Entropic regularization yields fast Sinkhorn solvers \cite{cuturi2013sinkhorn}. Schrodinger bridge (SB) and covariance-steering connect density control to linear systems with implementable controllers \cite{7160692,8264189,BAKOLAS201861}. Initial OMT studies drew inspiration from the SB problem \cite{7974883, marino2020optimal, chen2016relation, doi:10.1137/20M1320195} and cast OMT as transforming a probability density from a reference to a target configuration. OMT and SB methods offer distributional guarantees but typically require global optimization and dense computation \cite{benamou2000computational,MAL-073,cuturi2013sinkhorn}. Classical Voronoi Tessellations (CVT) coverage is scalable yet fragile under topology changes and noncooperation \cite{1284411,doi:10.1137/S0036144599352836}. Resilient consensus and containment focus on agreement or convex-hull invariance rather than target-set transport \cite{6481629,4118472}. References \cite{marino2020optimal, chen2016relation, doi:10.1137/20M1320195} analyze these connections in detail. While powerful, these approaches often rely on centralized optimization or PDE solvers and global models, limiting onboard, real-time resilience to agent dropouts—motivating decentralized alternative like ours. 

CVT and Lloyd-type controllers underpin distributed coverage \cite{doi:10.1137/S0036144599352836,1284411}. Power diagrams support heterogeneity and balance \cite{doi:10.1137/0216006}, and Coverage Path Planning (CPP) surveys document path-planning variants and constraints \cite{GALCERAN20131258}. Extensions treat non-convex or geodesic environments and heterogeneous sensing \cite{4739194,doi:10.1177/0278364913507324,5509696}. These methods iterate partitions or centroids and can degrade under communication losses. Our design instead fixes a sparse feed-forward depth, blends upstream references via scheduled convex weights, and reconfigures locally by pruning edges around uncooperative agents; avoiding repartitioning while retaining convex-hull containment. 

Byzantine-resilient consensus (e.g. W-MSR) ensures agreement among normal nodes despite adversaries \cite{6481629} and complements attack-detection or secure estimation in networked control \cite{6545301,MITRA2019108487}. Containment control guarantees followers remain within the convex hull of leaders for stationary or dynamic references \cite{4118472,https://doi.org/10.1002/rnc.3195,Thummalapeta26102023}. However, these strands typically target agreement or containment—not transport to a distributed target set—and do not prescribe assignment-free, layered feed-forward graphs with analytically scheduled convex weights. Our framework borrows the convex-hull principle while integrating resilient mentoring and edge-pruning to manage uncooperative agents without global reassignment. Interaction Networks and Graph Neural Networks (GNNs) enable learned multi-agent policies via message passing \cite{battaglia2016interaction, 4700287,pmlr-v100-tolstaya20a}. Such methods often require offline training, online inference, and changing graph supports. We instead use an analytically specified, fixed-depth feed-forward topology with time-scheduled barycentric weights—preserving interpretability and low compute while achieving coverage transport with resilience. 

UAS teams must transport and cover distributed targets despite intermittent links and agents that ignore coordination updates. Across OMT, SB, CVT or partitioning, resilient consensus or containment, and learning-based coordination, we find no assignment-free, decentralized method that (i) guarantees convex-hull containment; (ii) transports a team to a distributed target set; and (iii) handles uncooperative agents via local edge pruning with provable stability. This paper addresses this gap by extending the research done in \cite{11007873}, DNN transport to fault resilience with fixed feed-forward depth, forward weight scheduling, and mentor–mentee reconfiguration under uncooperation.

\subsection{Relation to prior work and Contributions}
This paper presents a DNN–based framework for decentralized transport of UAS in $\mathbb{R}^n$, enabling transitions from an initial formation to a final configuration that maximizes coverage of a distributed environmental target set. The proposed method introduces a structured DNN-driven communication architecture that abstracts the target set into a finite set of representative points, employs a purely forward training procedure with substantially lower computational cost than existing forward–backward learning algorithms, and eliminates convergence concerns. We further establish stability and asymptotic convergence guarantees, showing that all agents reliably reach the desired final formation representing the abstracted target set. Compared to both existing literature and the authors’ prior work \cite{11007873} (conference version), this paper offers the following new contributions:
\begin{enumerate}
    \item formally distinguishes cooperative and uncooperative agents and characterizes their roles within the multi-layer DNN;
    \item defines the Nominal DNN for fully cooperative teams and the Actual DNN that systematically incorporates uncooperative agents;
    \item demonstrates how fixed-position uncooperative agents induce communication-edge pruning and modified mentoring relations, thereby altering the effective information flow;
    \item analyzes how these structural modifications affect target coverage performance, including degradation introduced by uncooperative behavior;
    \item establishes new uniqueness and stability results under the pruned, time-varying convex communication topology; and
    \item presents expanded simulations with larger teams and diverse uncooperative patterns, providing quantitative assessments of resilience and performance.
\end{enumerate}

\subsection{Outline}

This paper is organized as follows. Section \ref{sec:problem} states the problem, vehicle model, and objectives for coverage-oriented transport. Section \ref{DNN Structure} specifies the feed-forward DNN: \ref{subsec:nominal-dnn} builds the nominal structure from the initial formation; \ref{subsec:actual-dnn} gives the actual deployment that isolates uncooperative agents via edge pruning while preserving convex mentoring. Section \ref{Target} abstracts the target set and computes final desired positions from target samples. Section \ref{Com_Weight}  derives initial/final barycentric weights and a smooth forward schedule for time-varying convex communication, and records the global desired set-points; uniqueness follows from the proposed construction. Section \ref{Multi-Agent Coverage Dynamics and Control} develops the coverage dynamics and tracking controller, and \ref{Stability} establishes stability and convergence conditions. Section \ref{Results} reports simulations under full cooperation and mixed cooperative–uncooperative scenarios; conclusions follow.

%
\IEEEpeerreviewmaketitle

\section{Problem Statement}\label{sec:problem}

We consider a team of $N$ quacopter UASs indexed by $\mathcal{V}=\{1,\ldots,N\}$. Every UAS $i\in \mathcal{V}$ is modeled by a nonlinear dynamics:

\begin{equation}\label{agenti}
    \mathbf{f}_i\left(\mathbf{x}_i,\mathbf{u}_i\right)=\mathbf{F}\left(\mathbf{x}_i\right)+\mathbf{G}\left(\mathbf{x}_i\right)\mathbf{u}_i
\end{equation}
where   $\mathbf{x}_i=\big[x_i, y_i, z_i, \dot{x}_i, \dot{y}_i, \dot{z}_i,\phi_i,\theta_i,\psi_i,\dot{\phi}_i,\dot{\theta}_i,\dot{\psi}_i,p_i,\dot{p}_i\big]^\top$ and  $\mathbf{u}_i=\begin{bmatrix}
u_{1,i}&u_{2,i}&u_{3,i}&u_{4,i} \end{bmatrix}^\top$ are the state and control of $i\in \mathcal{V}$. Translational positions are denoted by $x_i$, $y_i$ and $z_i$ and orientation angles $\phi_i$, $\theta_i$, and $\psi_i$ denote the roll, pitch, and yaw angles, respectively. The orientation (attitude) of quadcopter $i\in\mathcal{V}$ is represented using the Euler Z–Y–X (3–2–1) convention. For every agent $i\in \mathcal{V}$, we define $\mathbf{a}_i$, $\mathbf{r}_i$, $\mathbf{p}_i$ and $\mathbf{r}_{i,d}$ where $\mathbf{a}_i$ defines the initial position, $\mathbf{r}_i$ denotes the actual position, $\mathbf{p}_i$ defines the global desired position, and $\mathbf{r}_{i,d}$ represents the desired trajectory from the control output.

Our first objective is to determine $\mathbf{p}_i$ for every $i\in\mathcal{V}$ to maximize coverage of the distributed finite target set, $\mathcal{D}$. We develop a DNN-based strategy that organizes inter-UAS communication links based on the initial positions of the agents and maps $\mathcal{D}$ to $N$ representative points, $\{\mathbf{p}_1,\ldots,\mathbf{p}_N\}$ that serve as the desired positions for all the agents in $\mathcal{V}$. See Section \ref{Target} for details.

Our second objective is to safely coordinate the team of agents from $\Omega_{0}=\{\mathbf{a}_1,\ldots,\mathbf{a}_N\}$ to
$\Omega_{d}=\{\mathbf{p}_1,\ldots,\mathbf{p}_N\}$. For this purpose, we consider two scenarios. For scenario one, agent team $\mathcal{V}$ is divided into two groups, boundary leaders $\mathcal{V}_{0}$ and cooperative agents $\mathcal{V}_{c}$. Members of $\mathcal{V}_{0}$
 serve exclusively as leaders anchored to the boundary, providing stable references. Agents in $\mathcal{V}_{c}$ are role-flexible: depending on the current DNN formation, each can act as a leader for its mentees or as a mentee following an upstream mentor. Based on the DNN structure, we intend to establish inter-UAS communication links and specify communication weights such that best coverage of the target set $\mathcal{D}$ can be achieved.

 For scenario two, we divide the team of agents $\mathcal{V}$ into three groups: boundary leaders $\mathcal{V}_{0}$, cooperative agents $\mathcal{V}_{c}$ and uncooperative agents, $\mathcal{V}_{u}$. This setting models realistic contingencies in which any UAS may become uncooperative due to communication loss, malfunction, or mechanical faults and so on, while the mission still seeks maximal coverage of the target set $\mathcal{D}$, during the transfer $\Omega_{0}$ to $\Omega_{d}$. The key difference between the cooperative and uncooperative agents can be defined as:

\begin{equation}\label{rid}
    \mathbf{r}_{i,d}(t)=
    \begin{cases}
    \mathbf{p}_i&i\in \mathcal{V}_u\\
    \sum_{j\in \mathcal{I}_{i,j}}w_{i,j}(t)\mathbf{r}_j(t)&i\in \mathcal{V}_c 
    \end{cases}    
    .
\end{equation}
where, uncooperative UASs hold fixed desired positions at their assigned targets $\mathbf{p}_i$ and remain insensitive to network updates. Cooperative agents, in contrast, adapt their trajectories, $\mathbf{r}_{i,d}(t)$ via a time-varying convex combination of neighboring UASs, using nonnegative weights $w_{i,j}(t)$ that encode link availability and mentor–mentee relations. The DNN adapts links and weights to isolate $\mathcal{V}_{u}$, reassign mentors, and sustain coverage quality over $\mathcal{D}$ with graceful degradation. The details are followed in Section \ref{DNN Structure}. 

\section{DNN Structure}\label{DNN Structure}

This section specifies how the multi--UAS coverage problem is represented by a layered, feed–forward DNN. We first recall that every agent $i\in\mathcal{V}$ is represented by at least one neuron in the DNN. The neuron runs the agent’s differential dynamics and outputs its actual position $\mathbf{r}_i(t)$, which is then blended downstream. The DNN is therefore a graph $\mathcal{G}\left(\mathcal{V},\mathcal{E}\right)$ whose neurons are agents and whose directed edges indicate mentor--mentee relations with convex, time–varying weights. Two operational scenarios are considered: (i) boundary and cooperative agents only and (ii) boundary, cooperative, and uncooperative agents combined. \emph{Scenario~1} coincides with the \emph{Nominal DNN} synthesized from the initial formation; \emph{Scenario~2} minimally modifies that \emph{Nominal DNN} to accommodate clamped, uncooperative agents while preserving convex containment and forward propagation.

Based on the algorithm of operation, we distinguish between a \emph{Nominal DNN} and an \emph{Actual DNN}. The \emph{Nominal DNN} is synthesized by Algorithm \ref{euclid33} to represent standard DNN formation. The \emph{Actual DNN} is the deployed network and is presented in Sec.~\ref{subsec:actual-dnn} with \emph{Scenario~1} and \emph{Scenario~2}. \emph{Scenario~1} follows Algorithm~\ref{euclid33} as-is; while \emph{Scenario~2} introduces precise modifications to Algorithm \ref{euclid33} to incorporate uncooperative agents.

\begin{algorithm}
\caption{Construction of feed-forward network $\mathcal{G}(\mathcal{W},\mathcal{E})$}
\label{euclid33}
\begin{algorithmic}[1]
\State \textit{Input:} Get Initial agent positions $\{\mathbf{a}_i\}_{i=1}^N$
\State \textit{Output:} Obtain layer sets $\{\mathcal{W}_l\}_{l=0}^M$, edge set $\mathcal{E}$

\State Identify boundary agents $\mathcal{V}_B = \{b_1,\dots,b_{N_B}\}$.
\State Select core leader $b_{N_B+1}$ using \eqref{coreagent}; set $\mathcal{V}_0 = \mathcal{V}_B \cup \{b_{N_B+1}\}$.
\State Triangulate the convex polygon induced by $\mathcal{V}_0$ into
$\mathcal{S}_1,\dots,\mathcal{S}_{N_L}$.
\State Initialize open set $\mathcal{O} \gets \{\mathcal{S}_1,\dots,\mathcal{S}_{N_L}\}$,
$\mathcal{E} \gets \emptyset$, $\mathcal{W}_0 \gets \mathcal{V}_0$, and $\ell \gets 0$.

\While{$\mathcal{O} \neq \emptyset$}
  \State $\ell \gets \ell + 1$; $\mathcal{C} \gets \emptyset$; $\mathcal{W}_\ell \gets \mathcal{W}_{\ell-1}$
  \ForAll{$Q = \{i_1,\dots,i_{n+1}\} \in \mathcal{O}$}
    \State Choose candidate cooperative mentee agent $c_Q$
    \State inside simplex $Q$
           (if none exists, skip $Q$)
    \If{$c_Q$ exists}
      \State $\mathcal{W}_l \gets \mathcal{W}_{l-1} \cup \{c_Q\}$
      \State $\mathcal{N}_{c_Q} \gets Q$; add edges $(j,c_Q)$ for all $j \in \mathcal{N}_{c_Q}$ 
      \State to $\mathcal{E}$
      \State Append to $\mathcal{C}$ the simplexes generated by
      \State $\mathcal{EXP}(Q,c_Q)$
    \EndIf
  \EndFor
  \State $\mathcal{O} \gets \mathcal{C}$
\EndWhile

\State  Return $\mathcal{E}$ and $M$.
\end{algorithmic}
\end{algorithm}

\subsection{Nominal DNN}
\label{subsec:nominal-dnn}

The \emph{Nominal DNN} encodes the cooperative information flow implied by the initial formation. Its architecture is generated geometrically by Algorithm~\ref{euclid33}, using the positions
$\{\mathbf{a}_i\}_{i\in\mathcal{V}}$ of all agents similar to the one described in \cite{11007873}. DNN layers are indexed by $\mathcal{M}=\{0,\ldots,M\}$. We denote the set of agents first introduced as mentees at layer $l$ by $\mathcal{V}_l$ and the leaders by $\mathcal{V}_0$ . The neuron set of layer $l$ is defined cumulatively by
\begin{equation}
\mathcal{W}_l=\begin{cases}
\mathcal{V}_l & l\in\{0,M\},\\
\mathcal{V}_l \cup \mathcal{W}_{l-1} & l\in\mathcal{M}\setminus\{0,M\}
\end{cases}
\end{equation}
so that $\mathcal{W}_l$ collects all agents whose information can influence
mentees at layer $l$. By construction,
\[
\mathcal{V} = \bigcup_{l=0}^{M} \mathcal{V}_l,
\qquad
\mathcal{W}_0 \subset \mathcal{W}_1 \subset \cdots \subset \mathcal{W}_{M-1}
\]
and the edge set $\mathcal{E}\subset\mathcal{V}\times\mathcal{V}$ contains only
forward-directed connections from mentors to mentees.

For each agent $i$, the in-neighbor (mentor) set induced by
$\mathcal{E}$ is
\[
\mathcal{N}_i = \{\,j\in\mathcal{V} : (j,i)\in\mathcal{E}\,\}
\]
We distinguish two types of neurons in layer $l\ge 1$:
\[
\mathcal{I}_{i,l} =
\begin{cases}
\mathcal{N}_i, & i\in\mathcal{V}_l
= \mathcal{W}_l\setminus\mathcal{W}_{l-1} \quad \text{(new mentee)}\\[2pt]
\{i\}, & i\in\mathcal{W}_l\setminus\mathcal{V}_l
\quad \text{(carried-over mentor)}
\end{cases}
\]
Thus, a mentee at layer $l$ aggregates information only from upstream
neurons, while previously placed agents persist as identity neurons that relay
their states forward.

Unlike a standard neural network with static activations, each neuron $i$
implements the multicopter dynamics \eqref{agenti} and outputs the actual
position $\mathbf{r}_i(t)$. Given its mentor set $\mathcal{N}_i$, the desired
trajectory for neuron $i$ in the \emph{Nominal DNN} is
\begin{equation}\label{eq:rid_nominal}
\mathbf{r}_{i,d}(t)=
\begin{cases}
\mathbf{p}_i, \quad \quad \quad \quad \quad \quad \; i\in\mathcal{V}_0 \quad \text{(boundary agents)}\\[4pt]
\displaystyle\sum_{j\in\mathcal{N}_i} w_{i,j}(t)\,\mathbf{r}_j(t),
\; i\in\mathcal{V}\setminus\mathcal{V}_0 \hspace{2 pt} \text{(cooperative agents)}
\end{cases}
\end{equation}
where $\{w_{i,j}(t)\}_{j\in\mathcal{N}_i}$ are time-varying convex weights.
These weights are later designed so that each mentee remains inside the convex hull of its mentors, and the overall feed-forward structure propagates leader references through the network to realize the desired cooperative coverage behavior.

The terms explained by the Definitions below are applied by Algorithm \ref{euclid33} to obtain graph $\mathcal{G}\left(\mathcal{V},\mathcal{E}\right)$.

\begin{definition}\label{cores}
The position of the core agent is defined as the location of the non-boundary agent closest to the target center, expressed as
\begin{equation}\label{coreagent}
    b_{N_B+1}
    = \argmin_{ i \in (\mathcal{V} \setminus \mathcal{V}_{\text{0}})}
    \left\| \mathbf{a}_i - \mathbf{a}_{\text{tc}} \right\|_2
\end{equation}
where $\mathcal{V}$ denotes the set of all agents, $\mathcal{V}_{\text{0}}$ represents the boundary agents, and $\mathbf{a}_{\text{tc}}$ indicates the target center position.

\end{definition}

\begin{figure}[H]
\centering
\setlength{\tabcolsep}{0pt}     
\renewcommand{\arraystretch}{0} 

\resizebox{\linewidth}{!}{%
 
    \begin{tabular}{c}
      \includegraphics{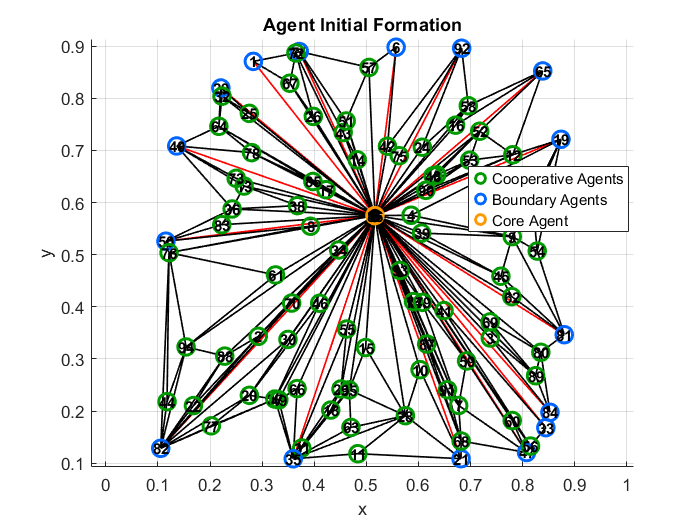}\\[0pt]
      
    \end{tabular}

} 

\caption{ Agent initial formation with cooperative agents in 2D motion space.}
\label{InitialFormation_Case1}
\end{figure}

Note that $\mathcal{V}_0=\{b_1,\dots,b_{N_B},b_{N_B+1}\}$ denote the leader set.
Figure~\ref{InitialFormation_Case1} illustrates a multi agent system with $N=95$ multicopters moving in a planar workspace ($n=2$). The formation is enclosed by a leading
polygon; its boundary agents are indexed by $\mathcal{V}_B$ with a total number of $N_B=16$. By Definition~\ref{cores}, the core leader is $b_{N_B+1}=b_{17}$, which
corresponds to agent $86$ (i.e., $b_{17}=86\in\mathcal{V}$). Hence, the leaders
are the boundary set together with the core leader, compactly:
\[
\mathcal{V}_0=\mathcal{V}_B\cup\{b_{N_B+1}\}.
\]

\begin{figure}[H]
\centering
\setlength{\tabcolsep}{0pt}     
\renewcommand{\arraystretch}{0} 

\resizebox{\linewidth}{!}{%
 
    \begin{tabular}{c}
      \includegraphics{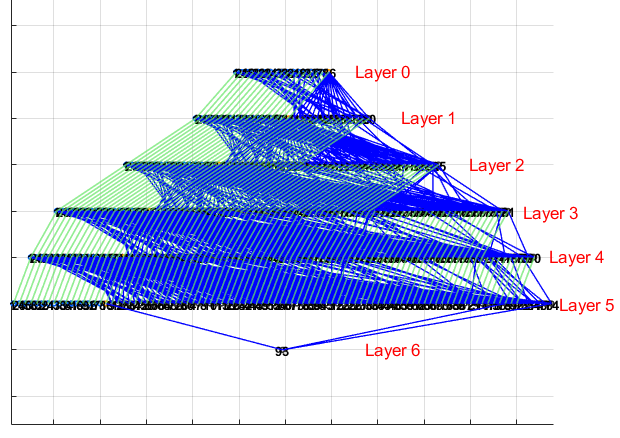}\\[0pt]
      
    \end{tabular}

} 
\caption{Feed-forward DNN structure based on the initial configuration of agents shown in Figure~\ref{InitialFormation_Case1}.}
\label{DNN_Case1}
\end{figure}

\begin{definition}
The \emph{leading polytope} is the convex hull generated by the positions of agents in $\mathcal{V}_B$.

\end{definition}

The leader set, $\mathcal{V}_0$ anchors the DNN and defines the leading polytope, i.e., the convex hull of $\mathcal{V}_B$. This polytope is partitioned into $N_L$ interior-disjoint 
$n$-simplices $\mathcal{S}_1,\ldots,\mathcal{S}_{N_L}$, each incident to the core leader. Here $N_L = 16$ (red lines); see Figure~\ref{InitialFormation_Case1}.

\begin{remark}
    
Algorithm~\ref{euclid33} then constructs a layered, feed-forward graph
$\mathcal{G}(\mathcal{V},\mathcal{E})$ as follows. At each iteration, every
simplex $Q = \{i_1,\ldots,i_{n+1}\}$ in the open set $\mathcal{O}$ is examined.
Among the agents whose initial positions lie in the simplex
$\Lambda(\mathbf{a}_{i_1},\ldots,\mathbf{a}_{i_{n+1}})$
, we select a \emph{cooperative mentee agent}
$c_{i_1,\ldots,i_{n+1}}$. This mentee is chosen
to be well-centered with respect to the simplex vertices and is connected to
its \emph{mentors} $\{i_1,\ldots,i_{n+1}\}$ via directed edges
$(i_k,c_{i_1,\ldots,i_{n+1}})$ for all $k$. Applying the expansion operator
$\mathcal{EXP}(\cdot)$ creates new simplexes that remain inside the leading
polytope, and these populate the updated open set $\mathcal{O}$. Iterating this
procedure eventually assigns each cooperative agent as a mentee exactly once,
thereby generating a feed-forward network \cite{11007873}.
\end{remark}

In summary, Algorithm~\ref{euclid33} produces a rigorously defined,
geometry-consistent communication graph $\mathcal{G}(\mathcal{V},\mathcal{E})$
that is directly interpreted as a Nominal feed-forward DNN for the UAS team.
\vspace{-10 pt}

\subsection{Actual DNN}
\label{subsec:actual-dnn}

The \textit{Actual} DNN is the deployed network that maintains the same forward-propagation law and convex constraints as the nominal model but operates under two distinct scenarios:

\begin{itemize}
\item \textbf{Scenario 1:} All non-boundary UASs are cooperative.
\item \textbf{Scenario 2:} At least one non-boundary UAS is uncooperative.
\end{itemize}

Under \emph{Scenario 1}, the \textit{Actual DNN}  is identical to the \textit{Nominal DNN}  and follows Algorithm \ref{euclid33} without any modification. Under \emph{Scenario 2}, the \textit{Actual DNN} remains time-invariant but differs structurally from the \textit{Nominal DNN}.
Therefore, this section mainly focuses on modifying the \textit{Nominal DNN} to derive the \textit{Actual DNN} in the presence of uncooperative non-boundary UASs.

For the second scenario, the group of UASs is defined as  
\begin{equation}
    \mathcal{V} = \mathcal{V}_0 \cup \mathcal{V}_c \cup \mathcal{V}_u,
\end{equation}
where $\mathcal{V}_0$, $\mathcal{V}_c$, and $\mathcal{V}_u$ denote the sets of boundary, cooperative, and uncooperative UASs, respectively. 
The key distinction from the cooperative-only case is architectural: uncooperative agents act as \emph{clamped sources}—they influence the network but do not adapt or propagate beyond the input layer. 

The layer-wise UAS sets are revised as follows:
\begin{equation}
    \hat{\mathcal{V}}_l =
    \begin{cases}
        \mathcal{V}_l \cup \mathcal{V}_u, & l = 0 \in \mathcal{M}, \\[4pt]
        \mathcal{V}_l \cup \mathcal{V}_c, & l \in \mathcal{M} \setminus \{0\}.
    \end{cases}
\end{equation}

Correspondingly, the neuron identification number sets are updated as
\begin{equation}\label{eq:case2-W}
    \mathcal{W}_l =
    \begin{cases}
        \hat{\mathcal{V}}_0, & l = 0, \\[4pt]
        \hat{\mathcal{V}}_l \cup \mathcal{W}_{l-1}, & l \in \mathcal{M} \setminus \{0, M\}, \\[4pt]
        \hat{\mathcal{V}}_l, & l = M.
    \end{cases}
\end{equation}
Each uncooperative UAS $i\in\mathcal{V}_u$ appears only in layer $0$ and provides fixed desired position $\mathbf{r}_{i,d}(t)=\mathbf{p}_i$, at any time $t$. This is because uncooperative UASs do not receive learned inputs from other UASs. By contrast, only cooperative UASs populate higher layers and update through \eqref{eq:rid_nominal} with time-varying convex weights $w_{i,j}(t)$. Consequently, \emph{Scenario~1} propagates adaptive influence from all non-boundary UASs, whereas \emph{Scenario~2} restricts adaptation to $\mathcal{V}_c$ and treats $\mathcal{V}_u$ as immutable boundary-like sources, preserving their constraints while enabling the cooperative subnetwork to reconfigure for target coverage. Algorithm \ref{alg:case2} shows the precise modifications to Algorithm \ref{euclid33} for \emph{Scenario~2}.

\begin{figure}[H]
\centering
\setlength{\tabcolsep}{0pt}     
\renewcommand{\arraystretch}{0} 

\resizebox{\linewidth}{!}{%
 
    \begin{tabular}{c}
      \includegraphics{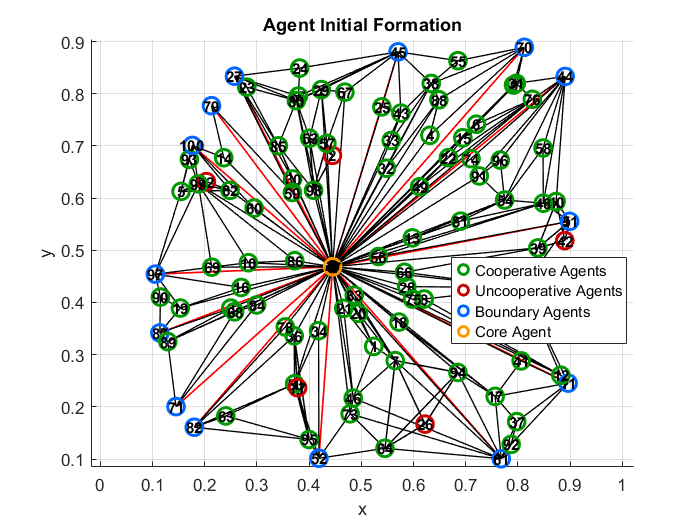}\\[0pt]
      
    \end{tabular}

} 

    \caption{An initial formation of an UAS team with $N=100$ UASs in a $2$-D motion space. The uncooperative UASs take part in the DNN structure but do not communicate with other UASs.}
\label{InitialFormation_Case2}
\end{figure}

\begin{figure}[H]
\centering
\setlength{\tabcolsep}{0pt}     
\renewcommand{\arraystretch}{0} 

\resizebox{1.02\linewidth}{!}{%

    \begin{tabular}{c}
      \includegraphics{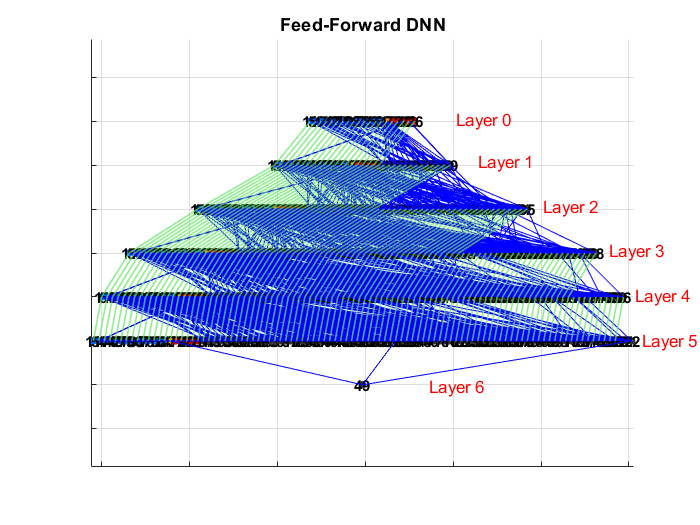}\\[0pt]
      
    \end{tabular}

} 

\caption{Feed forward DNN structure based on the initial configuration of UASs shown in Figure \ref{InitialFormation_Case2}.}
\label{DNN_Case2}
\end{figure}

\begin{algorithm}[ht]
  \caption{ Modifications to Nominal DNN}
  \label{alg:case2}
  \begin{algorithmic}[1]
    \State \textbf{Replace} $\mathcal{V}_0 \leftarrow \widehat{\mathcal{V}}_0=\mathcal{V}_0\cup\mathcal{V}_u$; set
           $\widehat{\mathcal{V}}_\ell\subset\mathcal{V}_c$ for $l\in\mathcal{M}\setminus\{0\}$ 
    \State \textbf{Form} $\mathcal{W}_l$ by \eqref{eq:case2-W}; \textbf{set} \eqref{eq:case2-clamped} for all $i\in\mathcal{V}_u$
    \State \textbf{Restrict} leaders for $l\in\mathcal{M}\setminus\{0\}$ and enforce \eqref{eq:case2-clamped} 
    \State \textbf{Constrain} edges per \eqref{eq:case2-edges}; no incoming edges to $\mathcal{V}_u$
    \State \textbf{Proceed} otherwise identically to Algorithm \ref{euclid33}
  \end{algorithmic}
\end{algorithm}

\begin{remark} 
Uncooperative UASs appear \emph{only} in layer~0 and are treated as fixed references:
\begin{equation}
  i\in \mathcal{V}_u:\quad
  \mathbf{r}_{i,d}(t)=\mathbf{p}_i,\qquad
  \mathcal{I}_{i,0}=\emptyset,\qquad
  \mathcal{N}_i=\emptyset
  \label{eq:case2-clamped}
\end{equation}
They do not receive mentors nor adapt their outputs; they serve as sources for $\mathcal{W}_l$ through edges in $\mathcal{E}$ emanating from $\mathcal{W}_{l-1}$.
\end{remark}

\begin{remark} 
For $l\in \mathcal{M}\setminus\{0\}$ and $i\in \widehat{\mathcal{V}}_l$,
\begin{equation}
  \mathcal{I}_{i,\ell}\;\subset\;\mathcal{W}_{\ell-1},
  \quad\text{and}\quad
  \mathbf{r}_{i,d}(t)=\sum_{j\in \mathcal{I}_{i,j}}
  w_{i,j}(t)\,\mathbf{r}_j(t)
  \label{Weight}
\end{equation}
with \[
w_{i,j}(t)\ge 0,\qquad \sum_{j\in\mathcal{N}_i} w_{i,j}(t)=1
\] enforced. Mentors in $\mathcal{W}_l$ include
boundary or uncooperative sources from $\widehat{\mathcal{V}}_0$
and/or cooperative UASs carried forward from lower layers.
\end{remark}

\begin{remark} 
We prohibit any incoming edge to uncooperative nodes:

\begin{equation}
  (j,i)\in\mathcal{E} \Rightarrow i\notin \mathcal{V}_u
  \label{eq:case2-edges}
\end{equation}
\end{remark}

Figure~\ref{InitialFormation_Case2} depicts the initial formation of a UAS team and the corresponding inter-UAS communication links obtained by following \emph{Scenario~2} and complying with Algorithm~\ref{alg:case2}.
 Boundary agents $\mathcal{V}_B$ (blue) and the core leader $b_{N_B+1}$ (orange) constitute layer~0 sources together with
uncooperative UASs $\mathcal{V}_u$ (red), which are clamped per
Algorithm \ref{alg:case2} and cooperative UASs $\mathcal{V}_c$ (green) populate deeper layers. Black edges visualize the communication graph and the mentor-mentee relations used to compute $\mathbf{r}_{i,d}(t)$ via convex blending. The figure also instantiates Defs.~1–2 and Remarks 1-4: (i) the leading polytope induced by $\mathcal{V}_B$ and the core leader; (ii) the UASs-in-simplex membership set $\mathcal{A}_{(\cdot)}$; (iii) the mentee-driven expansion operator $\mathcal{EXP}(\cdot)$ that refines each simplex around a selected mentee; and (iv) the open/closed families $(\mathcal{O},\mathcal{C})$ that track unexpanded versus expanded simplex cells. Together, these elements
illustrate the Scenario~2 deployment of the Actual DNN.

\begin{remark}
    In the following sections, we present the notations using \emph{Scenario~2}, i.e., we write
$\hat{\mathcal{V}}_0$ and $\hat{\mathcal{V}}_l$. When \emph{Scenario~1} is considered, these symbols
should be read as $\mathcal{V}_0$ and $\mathcal{V}_l$, respectively.

\end{remark}

\section{Abstract Representation of Target Zone} \label{Target}

We fix the core UAS $b_{N_B+1}\in\hat{\mathcal{V}}_0$ as the UAS whose initial position is closest to the center of the target zone, and we clamp its desired position to its initial position:
\[
\mathbf{p}_{b_{N_B+1}}=\mathbf{a}_{b_{N_B+1}}.
\]
For any cooperative UAS $i\in\hat{\mathcal{V}}_l$ with mentor set
$\mathcal{N}_i=\{i_1,\ldots,i_{n+1}\}$ (forming an $n$-simplex), we define the subset of target samples captured by its communication simplex as
\begin{equation}\label{Di}
\mathcal{D}_i=\Bigl\{\, r\in\mathcal{D}:\ \mathbf{d}_r\in
\Lambda\!\bigl(\mathbf{p}_{i_1},\ldots,\mathbf{p}_{i_{n+1}}\bigr)\Bigr\},
 i\in\hat{\mathcal{V}}_l, l\in\mathcal{M}\setminus\{0\}
\end{equation}
The final desired position of UAS $i$ is the average of the target samples within that simplex:
\begin{equation}\label{pi}
\mathbf{p}_i=\frac{1}{|\mathcal{D}_i|}\sum_{h\in\mathcal{D}_i}\mathbf{d}_h,
\qquad i\in\hat{\mathcal{V}}_l,\;\; l\in\mathcal{M}\setminus\{0\}
\end{equation}

\begin{assumption}
The target set $\mathcal{D}$ lies inside the leading convex polytope whose vertices are occupied by the boundary UASs.
\end{assumption}
\begin{assumption}
The target samples in $\mathcal{D}$ are stationary during the transport interval.
\end{assumption}

\section{Inter-UAS Communication Weights} \label{Com_Weight}
Let $\mathcal{V}_c$ denote the set of cooperative UASs, and let
$\mathcal{N}_i=\{i_1,\ldots,i_{n+1}\}$ be the $n{+}1$ in–neighbors of UAS
$i\in\mathcal{V}_c$. Because these neighbors form an $n$–simplex (i.e. are
affinely independent in $\mathbb{R}^n$), the augmented vertex matrix
\begin{equation}
\label{eq:aug-rank}
\mathrm{rank}\!\left(
\begin{bmatrix}
\mathbf{a}_{i_1} & \cdots & \mathbf{a}_{i_{n+1}}\\
1 & \cdots & 1
\end{bmatrix}
\right)=n{+}1,
\end{equation}
is full rank. Hence the initial position of each $i\in\mathcal{V}$ admits a (unique) barycentric representation with respect to its neighbors:
    \begin{subequations}\label{eq:init-bary}
        \begin{equation}
            \mathbf{a}_i=\sum_{j\in \mathcal{N}_i}\omega_{i,j}\mathbf{a}_j,\qquad \forall i\in \mathcal{V},
        \end{equation}
        \begin{equation}
            \sum_{j\in \mathcal{N}_i}\omega_{i,j}=1,\qquad \forall i\in \mathcal{V}.
        \end{equation}
    \end{subequations}
Solving \eqref{eq:init-bary} gives the (unique) initial communication weights
$\{\omega_{i,j}\}_{j\in\mathcal{N}_i}$:
\begin{equation}\label{eq:init-weights}
\begin{bmatrix}
\omega_{i,i_1}\\[-2pt]
\vdots\\[-2pt]
\omega_{i,i_{n+1}}
\end{bmatrix}
=
\begin{bmatrix}
\mathbf{a}_{i_1} & \cdots & \mathbf{a}_{i_{n+1}}\\
1 & \cdots & 1
\end{bmatrix}^{\!-1}
\begin{bmatrix}
\mathbf{a}_i\\
1
\end{bmatrix},\qquad i\in\mathcal{V}_c
\end{equation}
If $\mathbf{a}_i$ lies strictly inside the simplex of its neighbors, then all $\omega_{i,j}>0$ (and are nonnegative if $i$ lies on a face).

Let $\mathbf{p}_i$ denote the final desired positions obtained in Section~\ref{Target}. The
final barycentric weights of $i\in\mathcal{V}_c$ with respect to the same neighbor indices are
\begin{equation}\label{eq:final-weights}
\begin{bmatrix}
\varpi_{i,i_1}\\[-2pt]
\vdots\\[-2pt]
\varpi_{i,i_{n+1}}
\end{bmatrix}
=
\begin{bmatrix}
\mathbf{p}_{i_1} & \cdots & \mathbf{p}_{i_{n+1}}\\
1 & \cdots & 1
\end{bmatrix}^{\!-1}
\begin{bmatrix}
\mathbf{p}_i\\
1
\end{bmatrix},\qquad i\in\mathcal{V}_c
\end{equation}
which satisfy $\sum_{j\in\mathcal{N}_i}\varpi_{i,j}=1$ and are nonnegative (strictly positive if
$\mathbf{p}_i$ is in the interior of its neighbors' simplex).

Given the initial weights $\omega_{i,j}$ and final weights $\varpi_{i,j}$, each UAS defines its
time–varying communication weights via a smooth homotopy
\begin{equation}\label{eq:w-schedule}
w_{i,j}(t)=
\begin{cases}
\bigl(1-\beta(t)\bigr)\,\omega_{i,j}
+ \beta(t)\,\varpi_{i,j}, & t\in[t_0,t_f],\\[4pt]
\varpi_{i,j}, & t>t_f
\end{cases}
\end{equation}
where $\beta:[t_0,t_f]\!\to\![0,1]$ is an increasing $C^2$ quintic polynomial with $\beta(t_0)=0$ and $\beta(t_f)=1$. The minimum–jerk polynomial used is
\[
\beta(t)=10\tau^3-15\tau^4+6\tau^5,
\]
where 
\[
\tau=\frac{t-t_0}{t_f-t_0}\in[0,1].
\]

Let $N_0=|\mathcal{\hat{V}}_0|$ be the number of leaders and $N=|\mathcal{V}|$ the team size. The communication matrix,
$\mathbf{L}(t)=[L_{ij}(t)]\in\mathbb{R}^{N\times N}$ defined as,
\begin{equation}\label{eq:L-def}
L_{ij}(t)=
\begin{cases}
-1, & i=j,\\
w_{i,j}(t), & i\in \mathcal{V}\setminus\mathcal{\hat{V}}_0,\; j\in\mathcal{N}_i\\
0, & \text{otherwise.}
\end{cases}
\end{equation}
Rows corresponding to leaders $i\in\mathcal{\hat{V}}_0$,  have $-1$ on the diagonal and zeros elsewhere; rows of followers or cooperative UASs have
negative diagonal entries and nonnegative off–diagonal entries (convex weights), with the cooperative UAS row sum equal to zero.

\begin{definition}
The global desired position of each UAS $i\in\mathcal{V}$ is denoted $\mathbf{s}_i(t)$, with
$\mathbf{s}_i(t)=\mathbf{p}_i$ for all leaders $i\in\mathcal{\hat{V}}_0$.
\end{definition}

\begin{definition}
We aggregate the desired positions into
\[
\mathbf{S}(t)=\mathrm{vec}\!\left(\begin{bmatrix}
\mathbf{s}_1(t)&\cdots&\mathbf{s}_N(t)
\end{bmatrix}^{\!\top}\right)\in\mathbb{R}^{Nn\times 1},
\]
and define
\[
\mathbf{O}=\mathrm{vec}\!\left(\begin{bmatrix}
\mathbf{o}_1&\cdots&\mathbf{o}_N
\end{bmatrix}^{\!\top}\right)\in\mathbb{R}^{Nn\times 1},
\mathbf{o}_i=
\begin{cases}
\mathbf{p}_i,\quad i\in\mathcal{\hat{V}}_0\\
\mathbf{0}_{n\times 1}, i\notin\mathcal{\hat{V}}_0
\end{cases}
\]
\end{definition}

\begin{theorem}[Uniqueness of global desired set–points]
\label{thm:global-S}
For any $t$, the global desired positions $\{\mathbf{s}_i(t)\}_{i=1}^N$ are uniquely determined by
\begin{equation}\label{eq:global-S}
\bigl(\mathbf{I}_n\otimes \mathbf{L}(t)\bigr)\,\mathbf{S}(t)+\mathbf{O}=\mathbf{0}_{Nn\times 1}.
\end{equation}
\end{theorem}
{\renewcommand{\qedsymbol}{}
\begin{proof}
Index UASs so that leaders precede followers: let
$\mathcal{\hat{V}}_0=\{v_1,\ldots,v_{N_0}\}$ and
$\mathcal{V}_c=\{v_{N_0+1},\ldots,v_N\}$. With the permutation matrix $\mathbf{Q}$ that maps this
ordering, define
\[
\tilde{\mathbf{S}}(t)=(\mathbf{I}_n\otimes \mathbf{Q})\,\mathbf{S}(t)=
\begin{bmatrix}\tilde{\mathbf{S}}_L(t)\\ \tilde{\mathbf{S}}_I(t)\end{bmatrix},\quad
\tilde{\mathbf{O}}=(\mathbf{I}_n\otimes \mathbf{Q})\,\mathbf{O}=
\begin{bmatrix}\tilde{\mathbf{S}}_L\\ \mathbf{0}\end{bmatrix},
\]
and $\tilde{\mathbf{L}}=\mathbf{Q}\mathbf{L}\mathbf{Q}^\top=\begin{bmatrix}
-\mathbf{I}_{N_0} & \mathbf{0}\\
\tilde{\mathbf{L}}_{21} & \tilde{\mathbf{L}}_{22}
\end{bmatrix}$, where $\tilde{\mathbf{L}}_{21}\!\ge\!0$ and $\tilde{\mathbf{L}}_{22}$ has negative
diagonal and nonnegative off–diagonal entries. Equation \eqref{eq:global-S} becomes
\[
\left(\mathbf{I}_n\otimes
\begin{bmatrix}
-\mathbf{I}_{N_0} & \mathbf{0}\\
\tilde{\mathbf{L}}_{21} & \tilde{\mathbf{L}}_{22}
\end{bmatrix}\right)
\begin{bmatrix}\tilde{\mathbf{S}}_L\\ \tilde{\mathbf{S}}_I\end{bmatrix}
+
\begin{bmatrix}\tilde{\mathbf{S}}_L\\ \mathbf{0}\end{bmatrix}
=\mathbf{0}.
\]
The top block enforces $\tilde{\mathbf{S}}_L=\tilde{\mathbf{S}}_L$ (leaders clamped). The bottom block
yields
\[
\left(\mathbf{I}_n\otimes \tilde{\mathbf{L}}_{22}\right)\tilde{\mathbf{S}}_I
= -\left(\mathbf{I}_n\otimes \tilde{\mathbf{L}}_{21}\right)\tilde{\mathbf{S}}_L.
\]
By \cite{rastgoftar2021safe}, $\tilde{\mathbf{L}}_{22}$ is a nonsingular Hurwitz (M–matrix–type)
block; hence $\mathbf{I}_n\otimes\tilde{\mathbf{L}}_{22}$ is invertible and
\[
\tilde{\mathbf{S}}_I
= -\left(\mathbf{I}_n\otimes \tilde{\mathbf{L}}_{22}\right)^{-1}
  \left(\mathbf{I}_n\otimes \tilde{\mathbf{L}}_{21}\right)\tilde{\mathbf{S}}_L,
\]
which is unique. Transforming back with $\mathbf{Q}$ gives the stated result.
\end{proof}}

\section{UAS Team Coverage Dynamics and Control}\label{Multi-Agent Coverage Dynamics and Control}
This paper employs the input–state feedback linearization framework of \cite{rastgoftar2021safe} to design a trajectory–tracking controller for each multicopter similar to the paper \cite{11007873}. By defining the transformation
\[
\mathbf{x}_i \rightarrow \big(\mathbf{r}_i,\dot{\mathbf{r}}_i,\ddot{\mathbf{r}}_i,\dddot{\mathbf{r}}_i,\psi_i,\dot{\psi}_i\big),
\]
the original multicopter dynamics \eqref{agenti} can be converted to the following external (linearized) dynamics \cite{el2023quadcopter}:
\begin{subequations}\label{eq:extquad}
\begin{equation}\label{quaddynamicsext_rephrased}
\ddddot{\mathbf{r}}_i = \mathbf{v}_i,
\end{equation}
\begin{equation}
\ddot{\psi}_i = u_{\psi,i},
\end{equation}
\end{subequations}
where, following \cite{rastgoftar2022integration},
\begin{equation}
\mathbf{v}_i = \mathbf{M}_{1,i}\mathbf{u}_i + \mathbf{M}_{2,i},
\end{equation}
with
\begin{subequations}

\begin{equation}
\mathbf{M}_{1,i} =
\begin{bmatrix}
L_{\mathbf{g}_1}L_{\mathbf{F}}^{3}x_i & \cdots & L_{\mathbf{g}_4}L_{\mathbf{F}}^{3}x_i\\
L_{\mathbf{g}_1}L_{\mathbf{F}}^{3}y_i & \cdots & L_{\mathbf{g}_4}L_{\mathbf{F}}^{3}y_i\\
L_{\mathbf{g}_1}L_{\mathbf{F}}^{3}z_i & \cdots & L_{\mathbf{g}_4}L_{\mathbf{F}}^{3}z_i\\
L_{\mathbf{g}_1}L_{\mathbf{F}}\psi_i & \cdots & L_{\mathbf{g}_4}L_{\mathbf{F}}\psi_i
\end{bmatrix}.
\label{eq:M1i}
\end{equation}
where $\mathbf{M}_{1,i}\in\mathbb{R}^{4\times4}$.

\begin{equation}
\mathbf{M}_{2,i} =
\begin{bmatrix}
L_{\mathbf{F}}^{4}x_i &
L_{\mathbf{F}}^{4}y_i &
L_{\mathbf{F}}^{4}z_i &
L_{\mathbf{F}}^{2}\psi_i
\end{bmatrix}^T
\in \mathbb{R}^{4\times 1}.
\end{equation}
\end{subequations}

We assume the desired yaw is identically zero. Hence, choosing
\begin{equation}
u_{\psi,i} = -k_{5,i}\dot{\psi}_i - k_{6,i}\psi_i,\qquad k_{5,i}>0,\;k_{6,i}>0
\end{equation}
guarantees $\psi_i(t)=0$ for all $t$. Therefore, each multicopter $i\in\mathcal{V}$ is governed by \eqref{quaddynamicsext_rephrased}, which admits the state–space form
\begin{equation}\label{agentii_rephrased}
\begin{cases}
\dot{\mathbf{z}}_i = \mathbf{A}\mathbf{z}_i + \mathbf{B}\mathbf{v}_i,\\[2mm]
\mathbf{r}_i = \mathbf{C}\mathbf{z}_i,
\end{cases}
\qquad \forall i\in\mathcal{V},
\end{equation}
where $\mathbf{r}_i\in\mathbb{R}^{3\times 1}$ and $\mathbf{v}_i\in\mathbb{R}^{3\times 1}$ are, respectively, the output (position) and input vectors, and
\begin{subequations}
\begin{equation}
\mathbf{z}_i =
\begin{bmatrix}
\mathbf{r}_i^T & \dot{\mathbf{r}}_i^T & \ddot{\mathbf{r}}_i^T & \dddot{\mathbf{r}}_i^T
\end{bmatrix}^T
\in \mathbb{R}^{12\times 1},
\end{equation}
\begin{equation}
\mathbf{A} =
\begin{bmatrix}
\mathbf{0}_{9\times 3} & \mathbf{I}_{9}\\
\mathbf{0}_{3\times 3} & \mathbf{0}_{3\times 9}
\end{bmatrix}
\in \mathbb{R}^{12\times 12},
\end{equation}
\begin{equation}
\mathbf{B} =
\begin{bmatrix}
\mathbf{0}_{3\times 9} & \mathbf{I}_{3\times 3}
\end{bmatrix}^T
\in \mathbb{R}^{12\times 3},
\end{equation}
\begin{equation}
\mathbf{C} =
\begin{bmatrix}
\mathbf{I}_{3} & \mathbf{0}_{3\times 9}
\end{bmatrix}
\in \mathbb{R}^{3\times 12}.
\end{equation}
\end{subequations}

The coverage dynamics of the $l$-th UAS sub-team, whose UAS set is denoted by $\hat{\mathcal{V}}_l$, are described by
\begin{equation}\label{Layerldynamics}
    \begin{cases}
        \dot{\mathbf{X}}_l = \bar{\mathbf{A}}_l \mathbf{X}_l + \bar{\mathbf{B}}_l \mathbf{U}_l\\[1mm]
        \mathbf{Y}_l = \bar{\mathbf{C}}_l \mathbf{X}_l
    \end{cases}
    ,\qquad \forall\, l \in \mathcal{M},
\end{equation}
where
\[
\bar{\mathbf{A}}_l = \mathbf{I}_{N_l} \otimes \mathbf{A}, \qquad
\bar{\mathbf{B}}_l = \mathbf{I}_{N_l} \otimes \mathbf{B}, \qquad
\bar{\mathbf{C}}_l = \mathbf{I}_{N_l} \otimes \mathbf{C},
\]
and $\otimes$ denotes the Kronecker product. Let
\[
\hat{\mathcal{V}}_l = \{i_1,\dots,i_{N_l}\},
\]
then the stacked state, output, and input vectors are
\begin{subequations}
\begin{equation}
\mathbf{X}_l =
\begin{bmatrix}
\mathbf{z}_{i_1}^T & \cdots & \mathbf{z}_{i_{N_l}}^T
\end{bmatrix}^T,\qquad l\in\mathcal{M},
\end{equation}
\begin{equation}
\mathbf{Y}_l =
\begin{bmatrix}
\mathbf{r}_{i_1}^T & \cdots & \mathbf{r}_{i_{N_l}}^T
\end{bmatrix}^T,\qquad l\in\mathcal{M},
\end{equation}
\begin{equation}
\mathbf{U}_l =
\begin{bmatrix}
\mathbf{v}_{i_1}^T & \cdots & \mathbf{v}_{i_{N_l}}^T
\end{bmatrix}^T,\qquad l\in\mathcal{M}.
\end{equation}
\end{subequations}

For each UAS $i\in\mathcal{V}$, the virtual control input is selected as
\begin{equation}\label{vi}
    \mathbf{v}_i = -k_{1,i}\dddot{\mathbf{r}}_i
                   -k_{2,i}\ddot{\mathbf{r}}_i
                   -k_{3,i}\dot{\mathbf{r}}_i
                   +k_{4,i}\big(\mathbf{r}_{i,d} - \mathbf{r}_i\big),
    \qquad i\in\mathcal{V},
\end{equation}
where $\mathbf{r}_{i,d}(t)$ is the desired position defined in \eqref{rid}. Hence \eqref{vi} can be written in the compact form
\begin{equation}
    \mathbf{v}_i = -\mathbf{H}_i \mathbf{x}_i + \mathbf{G}_i \big(\mathbf{r}_{i,d} - \mathbf{C}\mathbf{x}_i\big),
\end{equation}
with
\[
\mathbf{H}_i = \begin{bmatrix}
\mathbf{0}_{n\times n} & \mathbf{H}_{2,i} & \cdots & \mathbf{H}_{4,i}
\end{bmatrix} \in \mathbb{R}^{3\times 9}, \qquad
\mathbf{G}_i \in \mathbb{R}^{3\times 3}.
\]
The feedback gain applied by UAS $i$ as
\begin{equation}
    \mathbf{K}_i = \mathbf{H}_i + \mathbf{G}_i \mathbf{C}.
\end{equation}
Then, for every multicopter $i \in \hat{\mathcal{V}}_l$, the closed-loop dynamics become
\begin{equation}\label{DynamicsofAgenti}
    \dot{\mathbf{z}}_i = \big(\mathbf{A} - \mathbf{B}\mathbf{K}_i\big)\mathbf{z}_i
                         + \mathbf{B}\mathbf{G}_i \mathbf{r}_{i,d}.
\end{equation}

\subsection{Equilibrium State}
The equilibrium of the closed–loop dynamics is attained when
\[
\big(\mathbf{A}-\mathbf{B}\mathbf{K}_i\big)\mathbf{z}_i + \mathbf{B}\mathbf{G}_i\mathbf{r}_{i,d} = \mathbf{0}.
\]
This condition can be written explicitly as
\begin{equation}
\begin{bmatrix}
\mathbf{0}_{3\times 3} & \mathbf{I}_3 & \mathbf{0}_{3\times 3} & \mathbf{0}_{3\times 3}\\
\mathbf{0}_{3\times 3} & \mathbf{0}_{3\times 3} & \mathbf{I}_3 & \mathbf{0}_{3\times 3}\\
\mathbf{0}_{3\times 3} & \mathbf{0}_{3\times 3} & \mathbf{0}_{3\times 3} & \mathbf{I}_3\\
-\mathbf{G}_i & -\mathbf{H}_{2,i} & -\mathbf{H}_{3,i} & -\mathbf{H}_{4,i}
\end{bmatrix}
\begin{bmatrix}
\mathbf{r}_i\\
\dot{\mathbf{r}}_i\\
\ddot{\mathbf{r}}_i\\
\dddot{\mathbf{r}}_i
\end{bmatrix}
+
\begin{bmatrix}
\mathbf{0}_{3\times 1}\\
\mathbf{0}_{3\times 1}\\
\mathbf{0}_{3\times 1}\\
\mathbf{G}_i\mathbf{r}_{i,d}
\end{bmatrix}
=
\begin{bmatrix}
\mathbf{0}_{3\times 1}\\
\mathbf{0}_{3\times 1}\\
\mathbf{0}_{3\times 1}\\
\mathbf{0}_{3\times 1}
\end{bmatrix}.
\end{equation}
Hence, the equilibrium is characterized by
\[
\mathbf{r}_{i,d} = \mathbf{r}_i,
\qquad
\frac{d^h \mathbf{r}_i}{dt^h} = \mathbf{0}, \; h=1,\dots,7.
\]
i.e., the position tracks the desired value and all higher–order derivatives vanish. This steady–state condition holds for all $t \ge t_f$, where the reference $\mathbf{s}_i(t)$ becomes constant, i.e., $\mathbf{s}_i(t)=\mathbf{p}_i$.

\subsection{Stability and Convergence Gurantee}\label{Stability}
Dynamics of layer $l\in\mathcal{M}$ in \eqref{Layerldynamics} can be rewritten as
\begin{equation}\label{CoverageDynamicsLayerl_rephrased}
    \dot{\mathbf{X}}_l = \big(\bar{\mathbf{A}}_l - \bar{\mathbf{B}}_l \bar{\mathbf{K}}_l\big)\mathbf{X}_l
    + \bar{\mathbf{B}}_l \bar{\mathbf{G}}_l \bar{\mathbf{V}}_l,\qquad \forall l\in\mathcal{M},
\end{equation}
where
\begin{subequations}
\begin{equation}
    \bar{\mathbf{K}}_l = \mathrm{diag}\big(\mathbf{K}_{i_1},\dots,\mathbf{K}_{i_{N_l}}\big), \qquad \forall l\in\mathcal{M},
\end{equation}
\begin{equation}
    \bar{\mathbf{G}}_l = \mathrm{diag}\big(\mathbf{G}_{i_1},\dots,\mathbf{G}_{i_{N_l}}\big), \qquad \forall l\in\mathcal{M},
\end{equation}
\begin{equation}
    \bar{\mathbf{V}}_l =
    \begin{bmatrix}
        \mathbf{r}_{i_1,d}^T & \cdots & \mathbf{r}_{i_{N_l},d}^T
    \end{bmatrix}^T,\qquad \forall l\in\mathcal{M}.
\end{equation}
\end{subequations}
The equilibrium of \eqref{CoverageDynamicsLayerl_rephrased} is attained if, for every $i\in\mathcal{V}$,
\[
\mathbf{r}_i = \mathbf{p}_i,
\qquad
\frac{d^h \mathbf{r}_i}{dt^h} = \mathbf{0},\; h=1,\dots,7.
\]

\begin{theorem}
Assume:
\begin{enumerate}
    \item For every $i\in\mathcal{V}$, the feedback gain $\mathbf{K}_i$ is selected so that $\mathbf{A}-\mathbf{B}\mathbf{K}_i$ is Hurwitz.
    \item For every $i\in\mathcal{V}_0$, the desired position $\mathbf{p}_i$ is constant.
\end{enumerate}
Then $\mathbf{r}_i(t)\to \mathbf{p}_i$ asymptotically for every $i\in\mathcal{V}$.
\end{theorem}

{\renewcommand{\qedsymbol}{}%
\begin{proof}
Given 1), $\mathbf{A}-\mathbf{B}\mathbf{K}_i$ is Hurwitz for all $i\in\mathcal{V}$. Since
\[
\bar{\mathbf{A}}_l - \bar{\mathbf{B}}_l \bar{\mathbf{K}}_l = \mathrm{diag}\big(\mathbf{A}-\mathbf{B}\mathbf{K}_{i_1},\dots,\mathbf{A}-\mathbf{B}\mathbf{K}_{i_{N_l}}\big),
\]
it follows that $\bar{\mathbf{A}}_l - \bar{\mathbf{B}}_l \bar{\mathbf{K}}_l$ is Hurwitz for every $l\in\mathcal{M}$. For layer $0$, item 2) implies $\mathbf{r}_{i,d}=\mathbf{p}_i$ is constant for all $i\in\hat{\mathcal{V}}_0$, hence $\mathbf{r}_i(t)\to \mathbf{p}_i$ as $t\to\infty$ and, in particular, $\mathbf{r}_j(t)$ is bounded for all $j\in\mathcal{\hat{V}}_0$. Because the inter-UAS communication weights are positive, the desired positions $\mathbf{r}_{i,d}$ of UASs in layer $1$ become bounded convex combinations of bounded leader states; thus \eqref{CoverageDynamicsLayerl_rephrased} is stable for $l=1$, and $\mathbf{r}_i(t)\to \mathbf{r}_{i,d}(t)$ for all $i\in\hat{\mathcal{V}}_l$. Proceeding inductively for $l=2,\dots,|\mathcal{M}|-1$, boundedness of layer $(l-1)$ positions implies bounded desired positions for layer $l$, so every UAS $i\in\hat{\mathcal{V}}_l$ tracks $\mathbf{s}_i(t)$ stably. Since $\mathbf{s}_i(t)=\mathbf{p}_i$ for all $t\ge t_f$, we conclude $\mathbf{r}_i(t)\to \mathbf{p}_i$ as $t\to\infty$ for every $i\in\mathcal{V}$.
\end{proof}}

\section{Simulation Results}\label{Results}

This section evaluates the performance of the proposed DNN-based UAS transport framework by comparing two distinct scenarios: (1) full cooperation with cooperative UASs, and (2) partial cooperation in the presence of uncooperative UASs. The convergence behavior of UASs toward a designated target zone is assessed using time-varying communication weights within a hierarchical DNN structure.


\subsection{DNN convergence with boundary and cooperative UASs only scenario}

The first scenario considers a team of $N=95$ UASs, consisting of 16 boundary UASs and 1 core UAS whose initial and final desired positions are fixed. The remaining 78 UASs are cooperative UASs participating in the DNN and dynamically adjusting their positions based on neighbor information. As shown in Fig. \ref{InitialFormation_Case1}, the UAS team begins from a 2D configuration, with triangulated connections forming the basis for hierarchical DNN construction. The resulting DNN structure is shown in Fig. \ref{DNN_Case1}.

\begin{figure}[!t]
\centering
\setlength{\tabcolsep}{0pt}     
\renewcommand{\arraystretch}{0} 

\resizebox{\linewidth}{!}{%
  
    \begin{tabular}{c}
      \includegraphics{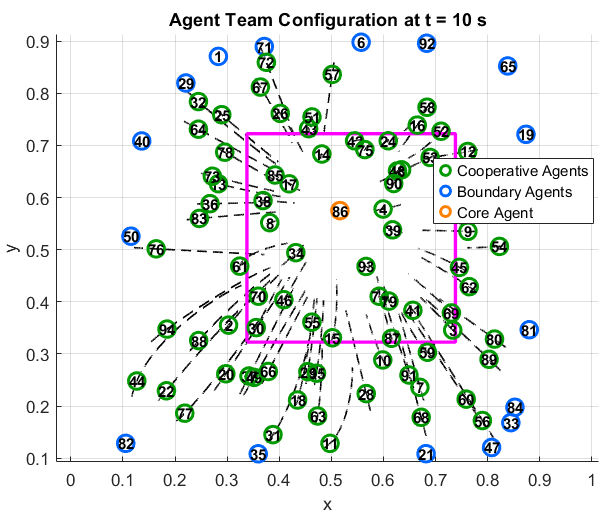}\\[0pt]
      
    \end{tabular}

} 

\caption{ UAS team transitional configuration at t = 10 s.}
\label{Regular_Converge_1}
\end{figure}

\begin{figure}[!t]
\centering
\setlength{\tabcolsep}{0pt}     
\renewcommand{\arraystretch}{0} 

\resizebox{\linewidth}{!}{%

    \begin{tabular}{c}
      \includegraphics{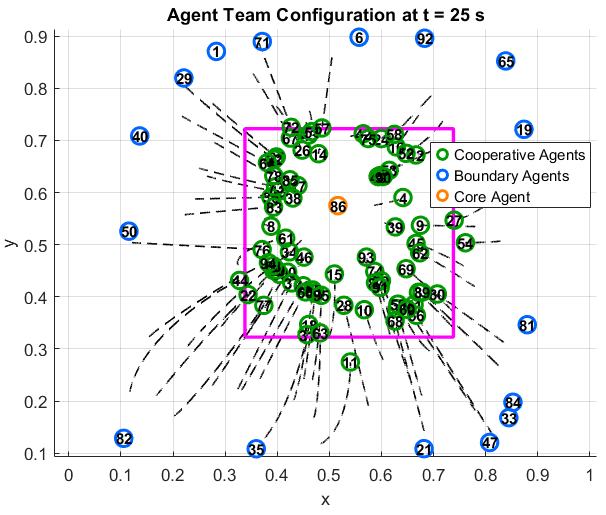}\\[0pt]
     
    \end{tabular}

} 

\caption{ All the cooperative UASs have converged towards the target zone at t = 25 s.}
\label{Regular_Converge_2}
\end{figure}

Figures \ref{Regular_Converge_1} and \ref{Regular_Converge_2} illustrate the convergence process over time. At t = 10 seconds, UASs are in motion toward their desired positions. By t = 25 seconds, all 78 cooperative UASs have successfully converged toward the target zone. A 10\% margin around the target zone boundary was included in determining successful convergence. Therefore, UAS $11$ falls into the converged UASs list. With all 78 UASs satisfying this criterion, the convergence rate in this scenario is 100\%; confirming the accuracy and efficiency of the DNN in guiding UASs toward coverage in a decentralized fashion.

\subsection{DNN convergence with boundary, cooperative and uncooperative UASs combined scenario}

\begin{figure}[!t]
\centering
\setlength{\tabcolsep}{0pt}     
\renewcommand{\arraystretch}{0} 

\resizebox{\linewidth}{!}{%

    \begin{tabular}{c}
      \includegraphics{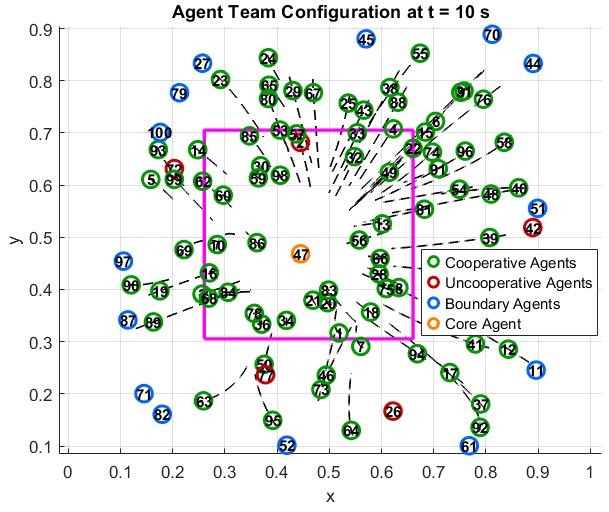}\\[0pt]
      
    \end{tabular}

} 

\caption{ UAS team transitional configuration at t = 10 s.}
\label{Stubborn_Converge_1}
\end{figure}

\begin{figure}[!t]
\centering
\setlength{\tabcolsep}{0pt}     
\renewcommand{\arraystretch}{0} 

\resizebox{\linewidth}{!}{%

    \begin{tabular}{c}
      \includegraphics{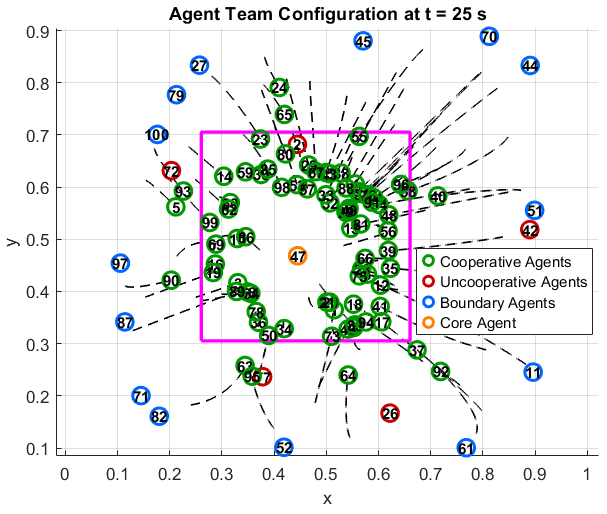}\\[0pt]
     
    \end{tabular}
  
} 

\caption{Except the uncooperative UASs and 7 of the cooperative UASs, all other cooperative UASs have converged successfully towards the target zone at t = 25 s (with a 10\% margin applied beyond border of the target zone for successful convergence).}
\label{Stubborn_Converge_2}
\end{figure}

In the second scenario, $N=100$ UASs are deployed, including 14 boundary UASs and 1 core UAS with fixed positions, and 5 uncooperative UASs that remain static throughout the simulation. It is observed from Figures \ref{Stubborn_Converge_1} and \ref{Stubborn_Converge_2} that one of the uncooperative UASs (UAS $2$) has both initial and final desired positions located inside the target zone and is excluded from convergence evaluation as its initial and final desired positions are fixed over time. The remaining 80 UASs are classified as cooperative UASs and are dynamically controlled by the DNN.

Figures \ref{Stubborn_Converge_1} and \ref{Stubborn_Converge_2} depict the system’s temporal evolution. Despite the presence of uncooperative UASs, the cooperative UASs exhibit effective convergence behavior. By t = 25 seconds, 73 of the 80 cooperative UASs reach the target zone (with a 10\% margin applied). This yields a convergence rate of 91.25\%, demonstrating strong fault resilience. The 7 unconverged cooperative UASs ($5, 24, 40, 64, 90, 92, 95$) are spatially distributed near the uncooperative UASs or peripheral regions, suggesting localized performance limitations.

These results confirm that the proposed DNN-based transport method ensures full convergence in cooperative settings and maintains high performance under partial cooperation. The architecture’s scalability, feedforward weight adaptation, and dynamic mentor-mentee assignments contribute to the robustness and adaptability of the multi-UAS coordination strategy.

\section{Conclusion}\label{Conclusion}

This paper introduces a fault–resilient DNN framework for decentralized UAS transport and coverage, unifying (i) layered, feed-forward inter-UAS topology, (ii) convex forward weight scheduling, and (iii) decentralized feedback with stability and uniqueness guarantees. The method explicitly addresses uncooperative UASs by deactivating position update throughout the simulation while preserving convex mentoring among cooperative UASs and computing global set-points consistent with leader references. The simulation-based evaluations confirm rapid coverage and strong resilience: under full cooperation, all coooperative UASs converge to the target zone using a 10\% margin. With five uncooperative UASs among $N{=}100$, 73 out of 80 cooperative UASs still converge (91.25\%), with the few misses localized near uncooperative or peripheral regions. These findings demonstrate graceful degradation, scalability, and low computational burden aligned with the design principles of the proposed architecture. Future work includes online detection and mitigation of adversarial behaviors, robustness to communication dropouts and disturbances, extension to dynamic targets and 3D constrained environments, and integration with formal safety monitors for guaranteed risk bounds.

 \section*{Acknowledgement}
  This work was partially supported by the National Science Foundation under Award Number 2133690.










\bibliographystyle{ieeetr}

\bibliography{citation} 

@article{benamou2000computational,
  title={A computational fluid mechanics solution to the Monge-Kantorovich mass transfer problem},
  author={Benamou, Jean-David and Brenier, Yann},
  journal={Numerische Mathematik},
  volume={84},
  number={3},
  pages={375--393},
  year={2000},
  publisher={Springer-Verlag Berlin/Heidelberg}
}

@article{MAL-073,
url = {http://dx.doi.org/10.1561/2200000073},
year = {2019},
volume = {11},
journal = {Foundations and Trends® in Machine Learning},
title = {Computational Optimal Transport: With Applications to Data Science},
doi = {10.1561/2200000073},
issn = {1935-8237},
number = {5-6},
pages = {355-607},
author = {Gabriel Peyré and Marco Cuturi}
}

@article{cuturi2013sinkhorn,
  title={Sinkhorn distances: Lightspeed computation of optimal transport},
  author={Cuturi, Marco},
  journal={Advances in neural information processing systems},
  volume={26},
  year={2013}
}

@ARTICLE{1284411,
  author={Cortes, J. and Martinez, S. and Karatas, T. and Bullo, F.},
  journal={IEEE Transactions on Robotics and Automation}, 
  title={Coverage control for mobile sensing networks}, 
  year={2004},
  volume={20},
  number={2},
  pages={243-255},
  doi={10.1109/TRA.2004.824698}}

@article{doi:10.1137/S0036144599352836,
author = {Du, Qiang and Faber, Vance and Gunzburger, Max},
title = {Centroidal Voronoi Tessellations: Applications and Algorithms},
journal = {SIAM Review},
volume = {41},
number = {4},
pages = {637-676},
year = {1999},
doi = {10.1137/S0036144599352836}
}

@ARTICLE{6481629,
  author={LeBlanc, Heath J. and Zhang, Haotian and Koutsoukos, Xenofon and Sundaram, Shreyas},
  journal={IEEE Journal on Selected Areas in Communications}, 
  title={Resilient Asymptotic Consensus in Robust Networks}, 
  year={2013},
  volume={31},
  number={4},
  pages={766-781},
  doi={10.1109/JSAC.2013.130413}}

@ARTICLE{4118472,
  author={Olfati-Saber, Reza and Fax, J. Alex and Murray, Richard M.},
  journal={Proceedings of the IEEE}, 
  title={Consensus and Cooperation in Networked Multi-Agent Systems}, 
  year={2007},
  volume={95},
  number={1},
  pages={215-233},
  doi={10.1109/JPROC.2006.887293}}

@INPROCEEDINGS{11007873,
  author={Rastgoftar, Hossein and Zahed, Muhammad J. H.},
  booktitle={2025 International Conference on Unmanned Aircraft Systems (ICUAS)}, 
  title={Deep Neural Network-Based UAS Transport}, 
  year={2025},
  volume={},
  number={},
  pages={1183-1189},
  doi={10.1109/ICUAS65942.2025.11007873}}

@article{GERBER2023102696,
title = {Optimal transport features for morphometric population analysis},
journal = {Medical Image Analysis},
volume = {84},
pages = {102696},
year = {2023},
issn = {1361-8415},
doi = {https://doi.org/10.1016/j.media.2022.102696},
author = {Samuel Gerber and Marc Niethammer and Ebrahim Ebrahim and Joseph Piven and Stephen R. Dager and Martin Styner and Stephen Aylward and Andinet Enquobahrie}
}

@article{wang2013linear,
  title={A linear optimal transportation framework for quantifying and visualizing variations in sets of images},
  author={Wang, Wei and Slep{\v{c}}ev, Dejan and Basu, Saurav and Ozolek, John A and Rohde, Gustavo K},
  journal={International journal of computer vision},
  volume={101},
  number={2},
  pages={254--269},
  year={2013},
  publisher={Springer}
}

@misc{lin20213d,
  title={3d brain tumor segmentation using a two-stage optimal mass transport algorithm. Sci Rep 11 (1)},
  author={Lin, WW and Juang, C and Yueh, MH and Huang, TM and Li, T and Wang, S and Yau, ST},
  year={2021}
}

@ARTICLE{7974883,
  author={Kolouri, Soheil and Park, Se Rim and Thorpe, Matthew and Slepcev, Dejan and Rohde, Gustavo K.},
  journal={IEEE Signal Processing Magazine}, 
  title={Optimal Mass Transport: Signal processing and machine-learning applications}, 
  year={2017},
  volume={34},
  number={4},
  pages={43-59},
  doi={10.1109/MSP.2017.2695801}}

@article{courty2017joint,
  title={Joint distribution optimal transportation for domain adaptation},
  author={Courty, Nicolas and Flamary, R{\'e}mi and Habrard, Amaury and Rakotomamonjy, Alain},
  journal={Advances in neural information processing systems},
  volume={30},
  year={2017}
}

@article{doi:10.1137/19M1261122,
author = {Stuart, Andrew M. and Wolfram, Marie-Therese},
title = {Inverse Optimal Transport},
journal = {SIAM Journal on Applied Mathematics},
volume = {80},
number = {1},
pages = {599-619},
year = {2020},
doi = {10.1137/19M1261122}
}

@article{WU2025107505,
title = {Optimal transport assisted full waveform inversion for multiparameter imaging of soft tissues in ultrasound computed tomography},
journal = {Ultrasonics},
volume = {147},
pages = {107505},
year = {2025},
issn = {0041-624X},
doi = {https://doi.org/10.1016/j.ultras.2024.107505},
author = {Xiaoqing Wu and Yubing Li and Chang Su and Panpan Li and Weijun Lin}

}

@article{ma2019optimal,
  title={Optimal mass transport based brain morphometry for patients with congenital hand deformities},
  author={Ma, Ming and Wang, Xu and Duan, Ye and Frey, Scott H and Gu, Xianfeng},
  journal={The Visual Computer},
  volume={35},
  number={9},
  pages={1311--1325},
  year={2019},
  publisher={Springer}
}

@ARTICLE{7053911,
  author={Su, Zhengyu and Wang, Yalin and Shi, Rui and Zeng, Wei and Sun, Jian and Luo, Feng and Gu, Xianfeng},
  journal={IEEE Transactions on Pattern Analysis and Machine Intelligence}, 
  title={Optimal Mass Transport for Shape Matching and Comparison}, 
  year={2015},
  volume={37},
  number={11},
  pages={2246-2259},
  doi={10.1109/TPAMI.2015.2408346}}

@article{shakib2020mass,
  title={Mass transfer evaluation in a multi-impeller extractor for reactive Mo (VI) extraction from aqueous Sulphate solution by utilizing coupling of acid and solvating Extractants},
  author={Shakib, Benyamin and Torab-Mostaedi, Meisam and Outokesh, Mohammad and Asadollahzadeh, Mehdi},
  journal={Heat and Mass Transfer},
  volume={56},
  number={6},
  pages={1995--2006},
  year={2020},
  publisher={Springer}
}

@article{song2023chest,
  title={Chest disease image classification based on spectral clustering algorithm},
  author={Song, Jiang and Gu, Yuan and Kumar, Ela},
  journal={Research Reports on Computer Science},
  pages={77--90},
  year={2023}
}

@ARTICLE{7160692,
  author={Chen, Yongxin and Georgiou, Tryphon T. and Pavon, Michele},
  journal={IEEE Transactions on Automatic Control}, 
  title={Optimal Steering of a Linear Stochastic System to a Final Probability Distribution, Part I}, 
  year={2016},
  volume={61},
  number={5},
  pages={1158-1169},
  doi={10.1109/TAC.2015.2457784}}

@INPROCEEDINGS{8264189,
  author={Goldshtein, Maxim and Tsiotras, Panagiotis},
  booktitle={2017 IEEE 56th Annual Conference on Decision and Control (CDC)}, 
  title={Finite-horizon covariance control of linear time-varying systems}, 
  year={2017},
  volume={},
  number={},
  pages={3606-3611},
  doi={10.1109/CDC.2017.8264189}}

@article{BAKOLAS201861,
title = {Finite-horizon covariance control for discrete-time stochastic linear systems subject to input constraints},
journal = {Automatica},
volume = {91},
pages = {61-68},
year = {2018},
issn = {0005-1098},
doi = {https://doi.org/10.1016/j.automatica.2018.01.029},
author = {Efstathios Bakolas}
}

@article{marino2020optimal,
  title={An optimal transport approach for the Schr{\"o}dinger bridge problem and convergence of Sinkhorn algorithm},
  author={Marino, Simone Di and Gerolin, Augusto},
  journal={Journal of Scientific Computing},
  volume={85},
  number={2},
  pages={27},
  year={2020},
  publisher={Springer}
}

@article{chen2016relation,
  title={On the relation between optimal transport and Schr{\"o}dinger bridges: A stochastic control viewpoint},
  author={Chen, Yongxin and Georgiou, Tryphon T and Pavon, Michele},
  journal={Journal of Optimization Theory and Applications},
  volume={169},
  number={2},
  pages={671--691},
  year={2016},
  publisher={Springer}
}

@article{doi:10.1137/20M1320195,
author = {Haasler, Isabel and Ringh, Axel and Chen, Yongxin and Karlsson, Johan},
title = {Multimarginal Optimal Transport with a Tree-Structured Cost and the Schrödinger Bridge Problem},
journal = {SIAM Journal on Control and Optimization},
volume = {59},
number = {4},
pages = {2428-2453},
year = {2021},
doi = {10.1137/20M1320195}
}

@article{doi:10.1137/0216006,
author = {Aurenhammer, F.},
title = {Power Diagrams: Properties, Algorithms and Applications},
journal = {SIAM Journal on Computing},
volume = {16},
number = {1},
pages = {78-96},
year = {1987},
doi = {10.1137/0216006}
}

@article{GALCERAN20131258,
title = {A survey on coverage path planning for robotics},
journal = {Robotics and Autonomous Systems},
volume = {61},
number = {12},
pages = {1258-1276},
year = {2013},
issn = {0921-8890},
doi = {https://doi.org/10.1016/j.robot.2013.09.004},
author = {Enric Galceran and Marc Carreras}
}

@INPROCEEDINGS{4739194,
  author={Pimenta, Luciano C. A. and Kumar, Vijay and Mesquita, Renato C. and Pereira, Guilherme A. S.},
  booktitle={2008 47th IEEE Conference on Decision and Control}, 
  title={Sensing and coverage for a network of heterogeneous robots}, 
  year={2008},
  volume={},
  number={},
  pages={3947-3952},
  doi={10.1109/CDC.2008.4739194}}

@article{doi:10.1177/0278364913507324,
author = {Subhrajit Bhattacharya and Robert Ghrist and Vijay Kumar},
title ={Multi-robot coverage and exploration on Riemannian manifolds with boundaries},
journal = {The International Journal of Robotics Research},
volume = {33},
number = {1},
pages = {113-137},
year = {2014},
doi = {10.1177/0278364913507324}
}

@INPROCEEDINGS{5509696,
  author={Breitenmoser, Andreas and Schwager, Mac and Metzger, Jean-Claude and Siegwart, Roland and Rus, Daniela},
  booktitle={2010 IEEE International Conference on Robotics and Automation}, 
  title={Voronoi coverage of non-convex environments with a group of networked robots}, 
  year={2010},
  volume={},
  number={},
  pages={4982-4989},
  doi={10.1109/ROBOT.2010.5509696}}

@ARTICLE{6545301,
  author={Pasqualetti, Fabio and Dörfler, Florian and Bullo, Francesco},
  journal={IEEE Transactions on Automatic Control}, 
  title={Attack Detection and Identification in Cyber-Physical Systems}, 
  year={2013},
  volume={58},
  number={11},
  pages={2715-2729},
  doi={10.1109/TAC.2013.2266831}}

@article{MITRA2019108487,
title = {Byzantine-resilient distributed observers for LTI systems},
journal = {Automatica},
volume = {108},
pages = {108487},
year = {2019},
issn = {0005-1098},
doi = {https://doi.org/10.1016/j.automatica.2019.06.039},
author = {Aritra Mitra and Shreyas Sundaram}
}

@article{https://doi.org/10.1002/rnc.3195,
author = {Li, Zhongkui and Duan, Zhisheng and Ren, Wei and Feng, Gang},
title = {Containment control of linear multi-agent systems with multiple leaders of bounded inputs using distributed continuous controllers},
journal = {International Journal of Robust and Nonlinear Control},
volume = {25},
number = {13},
pages = {2101-2121},
doi = {https://doi.org/10.1002/rnc.3195},
year = {2015}
}

@article{Thummalapeta26102023,
author = {Mourya Thummalapeta and Yen-Chen Liu},
title = {Survey of containment control in multi-agent systems: concepts, communication, dynamics, and controller design},
journal = {International Journal of Systems Science},
volume = {54},
number = {14},
pages = {2809--2835},
year = {2023},
publisher = {Taylor \& Francis},
doi = {10.1080/00207721.2023.2250041}
}

@article{battaglia2016interaction,
  title={Interaction networks for learning about objects, relations and physics},
  author={Battaglia, Peter and Pascanu, Razvan and Lai, Matthew and Jimenez Rezende, Danilo and others},
  journal={Advances in neural information processing systems},
  volume={29},
  year={2016}
}

@ARTICLE{4700287,
  author={Scarselli, Franco and Gori, Marco and Tsoi, Ah Chung and Hagenbuchner, Markus and Monfardini, Gabriele},
  journal={IEEE Transactions on Neural Networks}, 
  title={The Graph Neural Network Model}, 
  year={2009},
  volume={20},
  number={1},
  pages={61-80},
  doi={10.1109/TNN.2008.2005605}}

@InProceedings{pmlr-v100-tolstaya20a,
  title = 	 {Learning Decentralized Controllers for Robot Swarms with Graph Neural Networks},
  author =       {Tolstaya, Ekaterina and Gama, Fernando and Paulos, James and Pappas, George and Kumar, Vijay and Ribeiro, Alejandro},
  booktitle = 	 {Proceedings of the Conference on Robot Learning},
  pages = 	 {671--682},
  year = 	 {2020},
  editor = 	 {Kaelbling, Leslie Pack and Kragic, Danica and Sugiura, Komei},
  volume = 	 {100},
  series = 	 {Proceedings of Machine Learning Research},
  month = 	 {30 Oct--01 Nov},
  publisher =    {PMLR}
}

@article{rastgoftar2021safe,
  title={Safe affine transformation-based guidance of a large-scale multiquadcopter system},
  author={Rastgoftar, Hossein and Kolmanovsky, Ilya V},
  journal={IEEE Transactions on Control of Network Systems},
  volume={8},
  number={2},
  pages={640--653},
  year={2021},
  publisher={IEEE}
}

@article{rastgoftar2022integration,
  title={Integration of a* search and classic optimal control for safe planning of continuum deformation of a multiquadcopter system},
  author={Rastgoftar, Hossein},
  journal={IEEE Transactions on Aerospace and Electronic Systems},
  volume={58},
  number={5},
  pages={4119--4134},
  year={2022},
  publisher={IEEE}
}

@inproceedings{el2023quadcopter,
  title={Quadcopter tracking using euler-angle-free flatness-based control},
  author={El Asslouj, Aeris and Rastgoftar, Hossein},
  booktitle={2023 European Control Conference (ECC)},
  pages={1--6},
  year={2023},
  organization={IEEE}
}

\begin{IEEEbiography}[{\includegraphics[width=1in,height=1.25in,clip,keepaspectratio]{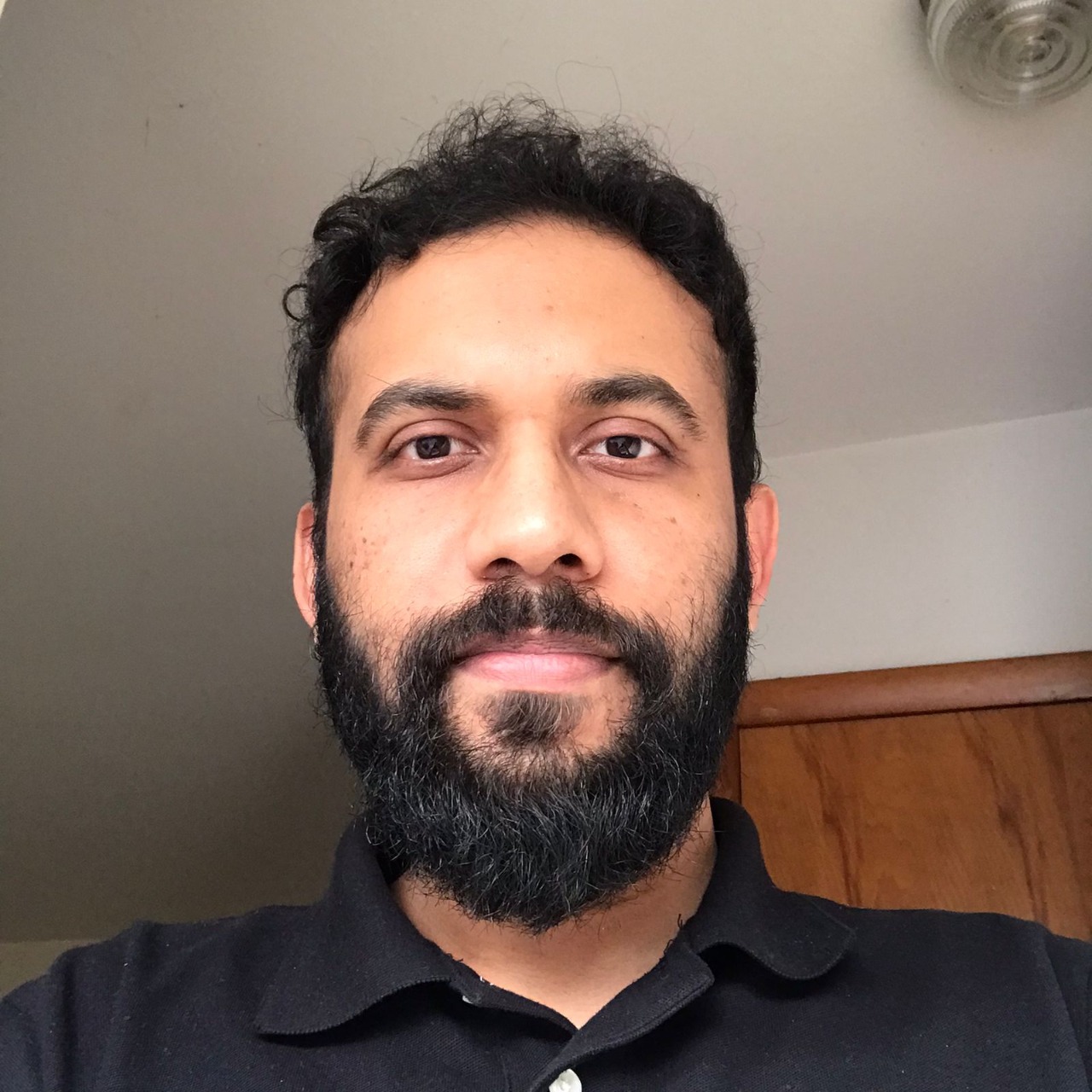}}]
{\textbf{Muhammad Junayed Hasan Zahed}} is a Graduate Research Assistant and a Ph.D. candidate in Aerospace Engineering at the University of Arizona. From 2022 to 2024, he served as a Graduate Research Assistant at The Pennsylvania State University. Previously, he was a Senior Faculty member (2020–2021) in the Department of Mechanical Engineering at the University of Creative Technology, Chittagong (UCTC). He holds a B.Sc. in Mechanical Engineering from Bangladesh University of Engineering and Technology (BUET), an M.S. in Mechanical Engineering from the University of Missouri–Kansas City (UMKC), and an M.S. in Aerospace Engineering from The Pennsylvania State University. His research interests span multi-agent systems, autonomous robots, control and developing algorithms for robotics applications. 

\end{IEEEbiography}
\vspace{-15cm} 
\begin{IEEEbiography}[{\includegraphics[width=1in,height=1.25in,clip,keepaspectratio]{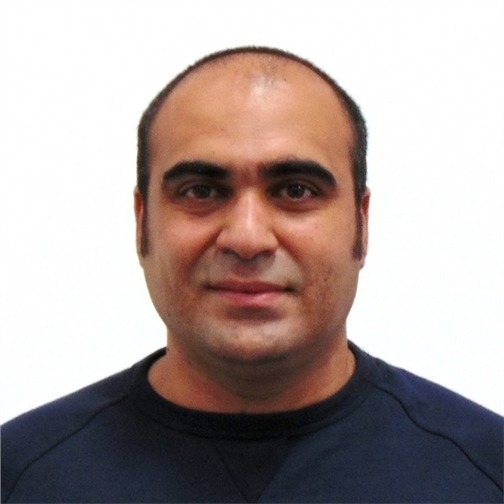}}]
{\textbf{Hossein Rastgoftar}} an Assistant Professor at the University of Arizona. Prior to this, he was an adjunct Assistant Professor at the University of Michigan from 2020 to 2021. He was also an Assistant Research Scientist (2017 to 2020) and a Postdoctoral Researcher (2015 to 2017) in the Aerospace Engineering Department at the University of Michigan Ann Arbor. He received the B.Sc. degree in mechanical engineering-thermo-fluids from Shiraz University, Shiraz, Iran, the M.S. degrees in mechanical systems and solid mechanics from Shiraz University and the University of Central Florida, Orlando, FL, USA, and the Ph.D. degree in mechanical engineering from Drexel University, Philadelphia, in 2015. 
\end{IEEEbiography}



%
\IEEEpeerreviewmaketitle

\end{document}